%% file: main.tex
\documentclass[acmsmall,screen]{acmart}
\usepackage{subcaption}
\usepackage{colortbl}
\usepackage[stable]{footmisc}
\usepackage{listings}
\usepackage{natbib}
\usepackage[ruled,linesnumbered]{algorithm2e}
\usepackage{syntax}
\usepackage{multirow}
\usepackage{simplebnf}
\usepackage{makecell}
\usepackage{array}
\usepackage{ragged2e}
\usepackage{threeparttable}
\usepackage{bbding}
\renewcommand\fbox{\fcolorbox{black}{gray!30}}

\lstdefinelanguage{Kotlin}{
  morekeywords={abstract, val, var, fun, if, else, while, for, in, return, open, class, interface, public, override},
  sensitive=true,
  morecomment=[l]{//},
  morecomment=[s]{/*}{*/},
  morestring=[b]",
  morestring=[s]{"""*}{*"""},
}
\lstdefinelanguage{Scala}{
  morekeywords={var, fun, if, else, while, for, in, return, open, class, interface, override, with, extends, c, abstract, def, return, public, default},
  sensitive=true,
  morecomment=[l]{//},
  morecomment=[s]{/*}{*/},
  morestring=[b]",
  morestring=[s]{"""*}{*"""},
}

\definecolor{codegreen}{rgb}{0,0.6,0}
\definecolor{codegray}{rgb}{0.5,0.5,0.5}
\definecolor{codepurple}{rgb}{0.58,0,0.82}
\definecolor{backcolour}{rgb}{0.95,0.95,0.92}

\begin{document}

\newcommand{\smallcapsbold}[1]{%
  \textsc{\textbf{\lowercase{#1}}}}

\newcommand{\kt}[1]{\href{https://youtrack.jetbrains.com/issue/KT-#1}{KT-#1}}
\newcommand{\scalaThree}[1]{\href{https://github.com/scala/scala3/issues/#1}{SCALA3-#1}}
\newcommand{\scalaTwo}[1]{\href{https://github.com/scala/bug/issues/#1}{SCALA2-#1}}
\newcommand{\groovy}[1]{\href{https://issues.apache.org/jira/browse/GROOVY-#1}{GROOVY-#1}}
\newcommand{\jdk}[1]{\href{https://bugs.openjdk.org/browse/JDK-#1?filter=allissues}{JDK-#1}}
\newcommand{\toolName}{\textsc{CrossLangFuzzer}}
\newcommand{\countKt}{10}
\newcommand{\countScalaTwo}{2}
\newcommand{\countScalaThree}{7}
\newcommand{\countGroovy}{4}
\newcommand{\countJava}{1}
\newcommand{\countAll}{24}

\newcommand{\ncheckmark}{\scalebox{0.85}[1]{$\times$}}

\newcommand{\dquote}[1]{``#1''}
\newcommand{\squote}[1]{`#1'}


\title{Finding Compiler Bugs through Cross-Language Code Generator and Differential Testing}

\author{Qiong Feng}
\email{qiongfeng@njust.edu.cn}
\affiliation{%
  \institution{Nanjing University of Science and Technology}
  \city{Nanjing}
  \country{China}
}

\author{Xiaotian Ma}
\email{xyzboom@njust.edu.cn}
\affiliation{%
  \institution{Nanjing University of Science and Technology}
  \city{Nanjing}
  \country{China}
}

\author{Ziyuan Feng}
\email{azumaseren@njust.edu.cn}
\affiliation{%
  \institution{Nanjing University of Science and Technology}
  \city{Nanjing}
  \country{China}
}

\author{Marat Akhin}
\email{marat.akhin@jetbrains.com}
\affiliation{%
  \institution{JetBrains}
  \city{Amsterdam}
  \country{The Netherlands}
}

\author{Wei Song}
\email{wsong@njust.edu.cn}
\affiliation{%
  \institution{Nanjing University of Science and Technology}
  \city{Nanjing}
  \country{China}
}

\author{Peng Liang}
\email{liangp@whu.edu.cn}
\affiliation{%
  \institution{Wuhan University}
  \city{Wuhan}
  \country{China}
}

\renewcommand{\shortauthors}{Feng et al.}


\input{abstract}




\maketitle

\input{intro}
\input{motivated}
\input{approach}
\input{experiment}

\input{results}
\input{discussion}
\input{related}

\input{conclusion}

\newpage

\bibliographystyle{ACM-Reference-Format}
\bibliography{ref}


\end{document}

%% file: abstract.tex
\begin{abstract}
Compilers play a central role in translating high-level code into executable programs, making their correctness essential for ensuring code safety and reliability. While extensive research has focused on verifying the correctness of compilers for single-language compilation, the correctness of cross-language compilation — which involves the interaction between two languages and their respective compilers — remains largely unexplored.

To fill this research gap, we propose \toolName{}, a novel framework that introduces a universal intermediate representation (IR) for JVM-based languages and automatically generates cross-language test programs with diverse type parameters and complex inheritance structures. After generating the initial IR, \toolName{} applies three mutation techniques — \textit{LangShuffler}, \textit{FunctionRemoval}, and \textit{TypeChanger} — to enhance program diversity. By evaluating both the original and mutated programs across multiple compiler versions, \toolName{} successfully uncovered \countKt{} confirmed bugs in the Kotlin compiler, \countGroovy{} confirmed bugs in the Groovy compiler, \countScalaThree{} confirmed bugs in the Scala 3 compiler, \countScalaTwo{} confirmed bugs in the Scala 2 compiler, and \countJava{} confirmed bug in the Java compiler. Among all mutators, \textit{TypeChanger} is the most effective, detecting 11 of the \countAll{} compiler bugs.

Furthermore, we analyze the symptoms and root causes of cross-compilation bugs, examining the respective responsibilities of language compilers when incorrect behavior occurs during cross-language compilation. To the best of our knowledge, this is the first work specifically focused on identifying and diagnosing compiler bugs in cross-language compilation scenarios. Our research helps to understand these challenges and contributes to improving compiler correctness in multi-language environments.

\end{abstract}

%% file: intro.tex
\section{Introduction}
Compilers play a central role in translating high-level code into executable programs, making their correctness crucial for ensuring code safety and reliability. Errors or inconsistencies in the compilation process can lead to unpredictable behavior, security vulnerabilities, and even system failures. Various approaches have been developed to generate diverse test programs and verify compiler correctness~\cite{chen2020survey}. For instance, most of generated-based compiler fuzzing techniques can generate programs with valid syntax to check whether they pass the compiler and produce the expected results, with invalid syntax to ensure they are rejected by the compiler, or programs that cause crashes to identify critical failures~\cite{csmith,YARPGen,ou2024mutators}. While these approaches~\cite{csmith,YARPGen,dewey2014language,dewey2015fuzzing,chaliasos2022finding,ou2024mutators} have been extensively studied for verifying single-language compilers, the correctness of compilers in cross-language systems remains largely unexplored.



Recent studies indicate that an increasing number of systems are implemented using multiple programming languages~\cite{yang2024multi}. For example, deep learning systems often combine C and Python~\cite{li2022polycruise}, while Linux systems start to use Rust in their C kernel~\cite{li2024empirical, panter2024rusty}. Furthermore, on the Java Virtual Machine (JVM), interoperability between Java, Kotlin, Groovy, and Scala allows them to coexist and evolve smoothly within a single project~\cite{ardito:2020ist, mateus:2020esem}. These four mainstream JVM languages (Java, Kotlin, Groovy, and Scala) frequently depend on one another at the source code level, posing challenges for compilers in resolving syntax and dependencies across different languages.



Consider the compilation process of a Kotlin-Java cross-language system as an example. The Kotlin compiler first resolves the Kotlin skeleton (class and method signatures) and passes them to the Java compiler. Given this skeleton, the Java compiler then compiles the Java source code into Java bytecode. Next, the Kotlin compiler processes the Java bytecode and compiles the Kotlin source code into Kotlin bytecode. As we can see, cross-language compilation involves two compilers and follows a cyclical process. Due to differences in type systems, language concepts, and compilation rules, cross-language compilation can easily introduce errors. However, little research has been conducted on these compilation processes or on testing the robustness of compilers in handling cross-language compilation~\cite{marr2016cross,garzella2020leveraging}.


Testing the correctness of compilers in handling cross-language compilation presents two significant challenges. First, generating cross-language interactive code is inherently difficult due to fundamental differences between programming languages. Each language has its own intermediate representation (IR), which is often incompatible with others. Furthermore, differences in syntax rules, type systems, and underlying programming concepts make it challenging to construct code that seamlessly integrates multiple languages. Ensuring that the generated code adheres to the constraints of each language while maintaining interoperability adds another layer of complexity. Second, even if cross-language code is generated successfully, determining whether it is semantically valid poses another challenge. Cross-language interactions often involve implicit conversions and type coercions that may behave differently depending on the compilation process. Without a clear specification of how different compilers should interpret such interactions, it becomes difficult to establish whether the generated code is functionally correct or whether subtle inconsistencies could lead to unexpected behavior or compilation failures.

To address the first challenge posed by differences in intermediate representations (IR) across programming languages, we propose \toolName{}, a novel framework designed to establish a universal IR for different JVM languages and systematically generate cross-language interaction code with various inheritance information and type parameters. By creating a standardized representation, \toolName{} ensures that the code can be seamlessly transformed and compiled across multiple languages within the JVM ecosystem. Furthermore, to enhance test program's diversity and improve bug detection, we introduce three novel mutators: \textit{LangShuffler} which randomly changes the code language, e.g., from Java to Kotlin; \textit{FunctionRemoval}, which eliminates functions in a class or an interface; and  \textit{TypeChanger}, which modifies type parameters to assess the robustness of type resolution mechanisms.

To address the second challenge—the difficulty of ensuring the correctness of generated cross-language code—we employ differential testing techniques. Since cross-language compilation involves compilers of different languages, we execute the generated code using two versions of one language's compiler: the latest release and a previous stable release, while keeping the other language's compiler fixed at its stable version. We then compare the compilation results to identify discrepancies and potential compiler miscompilations by minimizing the trigger program.

Using this approach, \toolName{} has generated diverse cross-language test programs with various type parameters and nontrivial inheritance structure. Within one month's bug hunting, \toolName{} successfully identified 10 confirmed bugs in the Kotlin compiler, 4 confirmed bugs in the Groovy compiler, 7 confirmed bugs in the Scala 3 compiler, 2 confirmed bugs in the Scala 2 compiler, and 1 confirmed bug in the Java compiler. These detected bugs exhibit diverse failure patterns and reveal fundamental weaknesses in cross-language compilation, highlighting the need for improved compiler robustness in multi-language ecosystems.
 

The \textbf{contributions} of this work are listed below:

\begin{enumerate}
    \item To the best of our knowledge, this is the first work to specifically target compiler bugs in JVM-based cross-language compilation. The designed IR in our framework enables the generation of interconnected class- and method-level programs across various languages, incorporating diverse type parameters and inheritance structures. Our framework demonstrates its effectiveness by uncovering \countAll{} bugs in JVM-based compilers.
    \item Based on the generated cross-language programs, we propose three novel mutation techniques—\textit{TypeChanger}, \textit{LangShuffler}, and \textit{FunctionRemoval}—which diversify test programs and have proven effective in uncovering cross-language compilation issues. 
    \item By analyzing the cross-language programs that trigger bugs in the cross-compilation process, we found that the user side (the language referencing a class or an interface from another language) tends to bear more responsibility for the bugs. We further discuss the responsibilities of different compilers in the cross-language compilation process.
\end{enumerate}

The rest of the paper is structured as follows: Section~\ref{sec:motivation} discusses the motivation of studying cross-language compiler bugs, Section~\ref{sec:approach} outlines the proposed approach, Section~\ref{sec:experiment} introduces the experiment setup and the RQs, Sections~\ref{sec:results} and~\ref{sec:discussion} present and discuss the results respectively, Section~\ref{sec:related} reviews the related work, and Section~\ref{sec:conclusion} concludes this work with future directions.


%% file: motivated.tex
\section{Motivating Examples}
\label{sec:motivation}

Kotlin is designed for seamless integration with Java; however, significant semantic differences between the two languages persist. For instance, Java retains the use of raw types to maintain compatibility with legacy versions that lack generics, while Kotlin does not. Additionally, there are subtle differences in class inheritance and method overriding strategies. These discrepancies create challenges in resolving Kotlin-Java type interactions within both compilers. The following two examples demonstrate how these differences are often inadequately addressed in compilers, leading to the introduction of compiler bugs.

\subsection{A Compiler Bug due to Discrepancies in the Use of Raw Types in Different Languages}
Kotlin is designed without raw types. A raw type can be understood as a type with unknown generic arguments in Kotlin. For example, a class declaration is \texttt{A0<T>} where \texttt{T} is a type parameter. When we declare \texttt{A0<String>}, it means that \texttt{String} is used as the type argument. However, when we declare \texttt{A0} without specifying a type argument, this represents a raw type in Java. In Kotlin, raw types like \texttt{A0} are not allowed. When interacting with Java, a raw type in Java is treated as a wildcard type \texttt{A0<*>} (similar to \texttt{A0<?>} in Java) by the Kotlin compiler~\cite{kotlinSpec} (for simplicity, we have omitted certain details, such as nullability, which are not related to the bug shown below). 

\begin{figure}
\centering
\begin{minipage}[c]{0.6\textwidth}
\begin{lstlisting}[language=Java]
// Java Files
public interface A<T> {
    void foo(List<T> list);
}
public abstract class B implements A<String> {
    @Override
    public final void foo(List list) {}
}
public class C extends B implements A<String> {}
// Kotlin File
class X: C() //compiler error
\end{lstlisting}
\end{minipage}
\caption{\kt{55822}: Kotlin compiler rejects a valid program with a raw type in Java class parameter}
\label{fig:example1}
\end{figure}

Figure~\ref{fig:example1} shows a Kotlin compiler bug that incorrectly handles the raw type in Java code. We can see that in \texttt{class B}, method \texttt{foo} overrides the method \texttt{A.foo} with a raw type. \texttt{Class C} extends \texttt{class B} and implements interface \texttt{A} with type argument \texttt{String}. When inheriting \texttt{class C} with a Kotlin \texttt{class X}, the Kotlin compiler rejects this valid program and reports ``class `\texttt{X}' is not abstract and does not implement abstract base class member \texttt{public abstract fun foo(list: (Mutable)List<String!>!): Unit}''. If we rewrite \texttt{class X} in Java source code, the Java compiler can pass the code with no problems.

As we can see, the Kotlin compiler fails to recognize that \texttt{B.foo} overrides \texttt{A.foo} because it interprets the raw type \texttt{List} as \texttt{List<*>} (this is a simplified translation; in actual Kotlin-Java interactions, the raw type \texttt{List} is treated as a flexible type \texttt{List<*>..MutableList<*>?} in Kotlin). Consequently, the compiler incorrectly assumes that \texttt{B.foo} does not override \texttt{A.foo}, as the wildcard \texttt{*} flexible type does not align with the generic type \texttt{T} in the interface. If the method in \texttt{class B} uses a generic type, or if \texttt{class C} does not explicitly declare the dependency on \texttt{A<String>}, the program works as expected. The Kotlin development team has addressed this bug (see at \kt{55822}).

\subsection{Compiler Bug due to Discrepancies in Handling Method Overrides in Interfaces and Parent Classes in Different Languages}
\label{subsec:motivation-2}

Figure~\ref{fig:example2} shows another existing bug in the Kotlin compiler. \texttt{Class C} extends \texttt{class A} and implements \texttt{interface IB}. If this code is written in Java, the method \texttt{foo} in \texttt{class C} is inherited from \texttt{class A}, meaning that in Java, the priority of a method from the parent class takes precedence over that from the parent interface. However, the same code fails to compile in Kotlin, as Kotlin treats the priorities of these methods as equal, resulting in a conflict. 


Inspired by this bug, we want to know whether discrepancies in method overrides can affect cross-language compilation. Figure~\ref{fig:example2-1} illustrates a program generated by our \toolName{} framework, involving Kotlin and Java inheritance, that triggers the bug identified as \kt{74151}. The Java \texttt{class A2} extends the Kotlin \texttt{class A1} and implements \texttt{I1} (see Line 16). As mentioned earlier, \texttt{class A2} should compile without issues since it is written in Java, and \texttt{class A2} uses \texttt{A1::func} as its implementation. However, when the Kotlin \texttt{class A3} extends \texttt{A2} and implements \texttt{I1} and \texttt{I0}, one of the latest Kotlin compiler versions rejects this valid program and reports \dquote{\texttt{Class \squote{A3}} must override public open fun \texttt{func}}, while the other version accepts it. This inconsistency between two versions of the Kotlin compiler proves that this is a compiler bug. In this case, if \texttt{class A2} were written in Kotlin, it would not compile because \texttt{A1::func} conflicts with \texttt{I1::func}. However, with the Java \texttt{class A2} acting as a buffer, the bug propagates through this Kotlin-Java-Kotlin inheritance chain. This example demonstrates that discrepancies in method overrides among different languages can cause cross-language compilation issues.

\begin{figure}
\centering
\begin{minipage}[c]{0.3\textwidth}
\begin{lstlisting}[language=Kotlin]
//All Kotlin Code 
open class A {
    fun foo() {}
}
interface IB {
    fun foo() {}
}
class C: A(), IB
\end{lstlisting}
\end{minipage}
\caption{\kt{10195}: Kotlin compiler rejects this code when the interface and the parent class of a subclass contain a method with the same signature}
\label{fig:example2}
\end{figure}

\begin{figure}
\centering
\begin{minipage}[c]{0.6\textwidth}
\begin{lstlisting}[language=Java]
// Kotlin File
interface I0 {
    fun func(arg0: ArrayList<Any>)
}
interface I1: I0 {
    override fun func(arg0: ArrayList<Any>) {
        println("I1")
    }
}
abstract class A1: I0 {
    override fun func(arg0: ArrayList<Any>) {
        println("A1")
    }
}
// Java File
public class A2<T> extends A1 implements I1 {
}
//Kotlin File
class A3: A2<Any>(), I1, I0 // could not compile

\end{lstlisting}
\end{minipage}
\caption{\kt{74151}: One Kotlin compiler passes this program while the other does not when a Kotlin class’s parent class in Java and interface in Kotlin contain methods with the same signature}
\label{fig:example2-1}
\end{figure}

\subsection{Observations from the Above Examples}

The examples provided above highlight how improper handling of syntax and design differences between Kotlin and Java can lead to various compiler bugs. Not only do differences exist in inheritance and generics, but discrepancies in nullability and declaration-site variance~\cite{kotlinSpec} can also introduce significant complexity into the interactions between the two languages. Given the complexities and discrepancies involved, we conjecture that there are likely still undiscovered issues related to Kotlin-Java interactions that may affect compiler behavior in subtle ways. Identifying and addressing these issues is essential for evaluating the robustness of Kotlin and Java compilers and ensuring smoother interactions between Kotlin and Java. 

Furthermore, languages like Scala and Groovy, which are also designed to interact seamlessly with Java, could face similar challenges due to design differences. Therefore, systematic testing of cross-language interactions is crucial, as it may uncover additional compiler-related issues and reveal further opportunities to enhance JVM-based language interoperability and reliability.

%% file: approach.tex
\section{Approach}
\label{sec:approach}

\begin{figure}
    \centering
    \includegraphics[width=0.85\linewidth]{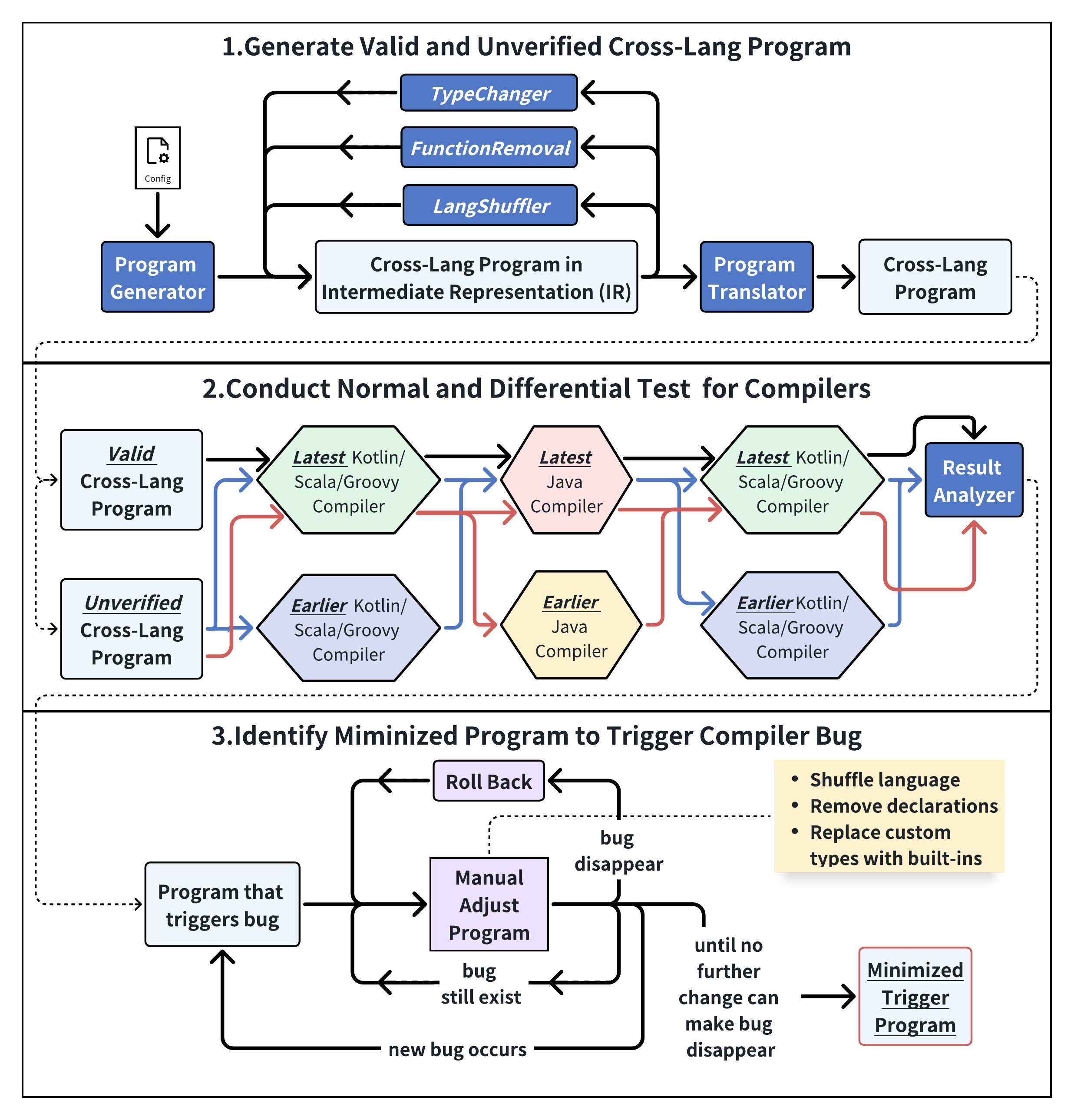}
    \caption{The \toolName{} Framework}
    \label{fig:framework}
\end{figure}

As shown in Figure~\ref{fig:framework}, our \toolName{} framework consists of three stages: \textit{Program Generation}, \textit{Compiler Testing}, and \textit{Trigger Program Minimization}. The core module of our approach is \textit{Program Generator}, which produces cross-language programs in an intermediate representation (IR). The generator can create cross-language programs with diverse semantic and syntactic characteristics. These programs follow the definition of IR, and we assert that they are semantically and syntactically valid and the compiler should pass them.

The generated IR programs can be further mutated by our three designed mutators. During mutation, we employ specific mechanisms (which will be explained in Section 3.2) to prevent the generated programs from deviating excessively and producing a large number of invalid programs. However, the validity of these programs has not been verified. For these programs, we use differential testing, asserting that different compiler implementations should produce the same compilation results. If the assertion fails, we successfully identify a program that triggers a compiler bug (In the following text, this program is referred to as a trigger program).

After identifying a trigger program, we minimize it before reporting the bug to the development team. Minimizing can remove all trigger program parts unrelated to bugs in order to help developer reproduce the bug and quickly identify its root cause without sifting through unnecessary code. During the minimization process, we strive to preserve the program's semantics while removing unrelated code. If a change makes the bug disappear, we roll it back. However, if it triggers a new bug, we add it to the set of trigger programs and start a new round of minimization. Once the program can no longer be further minimized, we report the bug.


\input{approach-ir}

\input{approach-genprog}
\input{approach-test}
\input{approach-manual}

%% file: approach-ir.tex
\subsection{Intermediate Representation Definition}
\label{subsec:ir} 

Figure~\ref{fig:ir} illustrates the syntax and types in the Intermediate Representation (IR) of cross-language program. A program ({\ttfamily prog}) in the IR consists of a sequence of top-level declarations, denoted as $\overline{d}$. A top-level declaration ({\ttfamily d}) can represent either a class or an interface with a language tag ({\ttfamily lang}) that indicates the programming language used in the generated source code. Each class or interface ({\ttfamily c}) contains a series of methods, denoted as $\overline{f}$. Classes and interfaces are defined using non-access modifiers, such as open, final, and abstract. For simplicity, the modifier details are not shown in Figure~\ref{fig:ir}.

The notation {\ttfamily class C extends $\overline{st}$} indicates that a class or an interface can extend a parent class or another interface. Additionally, a class can extend a parent class while implementing multiple interfaces simultaneously. A class may also include a series of member methods ($\overline{f}$). Member methods can also be defined with different modifiers, such as normal, abstract, or final. These modifiers result in three types of functions inherited from the parent class: (1) ``can override'', (2) ``must override'', and (3) ``cannot override''. For functions that must be overridden, an override function in the child class is created, denoted as ``{\ttfamily override} $\overline{f}$''. For functions that can be overridden, we randomly decide whether to create an override or not. 

To support the detection of generic type discrepancies between different languages, this IR supports parameterized declarations by introducing type parameters in class or interface declarations, denoted as ``{\ttfamily class C $\tau$}''. Accordingly, the supertype $st$ is denoted as ``\underline{c}.C $\overline{\tau}$'' to support type parameters in supertypes. Currently this IR does not support expression and access modifiers as we mainly focus on how compiler processes inheritance and generics. There are several reasons why we introduced a new IR instead of using JVM bytecode. First, JVM bytecode lacks key information we need. For example, we need to record all the overridden methods of a given method, but this information is not available in the bytecode. Second, bytecode includes details irrelevant to our goals, such as synthetic bridge methods for generics, which don’t affect source-level semantics. Third, bytecode can misrepresent source code behavior. For instance, due to type erasure, methods that don't override at the source code level may appear to do so in bytecode.

\begin{figure}
\centering
\begin{bnf}(prod-delim={;;})[ column{1}={font=\sffamily}, column{2}={font=\ttfamily},  column{3}={font=\ttfamily}, column{4}={font=\ttfamily}, column{5}={font=\ttfamily}, column{6}={font=\ttfamily}, column{7}={font=\ttfamily}]
<prog $\in$ Program> ::= $\overline{d}$ ;;
<d $\in$ TopDecleration> ::= c lang // \underline{c}  lang ;;
<c $\in$ ClassOrInterface> ::= class C extends $\overline{st}$ contains $\overline{f}$ ;; 
<\underline{c} $\in$ ParamterizedClassOrInterface> ::= class C $\overline{\tau}$ extends $\overline{st}$ contains $\overline{f}$ ;;
<f $\in$ MemberMethod> ::= method m $\overline{param}$ $\colon$$t$ override $\overline{f}$ ;; 
<param $\in$ Parameter> ::= p$\colon$t ;; 
<C $\in$ ClassName> ::= the set of Class name ;; 
<m $\in$ methodName> ::= the set of method name ;; 
<p $\in$ parameterName> ::= the set of parameter name ;; 
<$\tau$ $\in$ TypeParameter> ::= the set of type parameter ;;
<st $\in$ AvailableSuperType> ::= c.C // \underline{c}.C $\overline{\tau}$;;
<t $\in$ Type> ::= st // $\tau$ ;;
<lang $\in$ Language> ::= the language of the declaration ;;
\end{bnf}
\caption{The syntax and types in the IR of cross-language program}
\label{fig:ir}
\end{figure}

%% file: approach-genprog.tex
\subsection{Generate Valid and Unverified Cross-Language Program}
\label{subsec:gen-program-alog}

\input{algo1}
\input{algo2} 

Algorithm~\ref{algo:overall} presents the overall process for generating a cross-language program. First, a class or interface name is created with a random language tag. For example, a Kotlin class such as \texttt{class A} might be generated at this stage. Next, type parameters are generated for the class or interface, e.g., \texttt{T} in \texttt{A<T>}. Currently, raw types and wildcards are not supported. Then, we randomly select this new generated class/interface's supertypes from the pool of previously generated classes or interfaces and generate the member methods of these supertypes. When selecting type arguments for a supertype, we choose from the new generated class's type parameters. The member methods in the supertypes are generated with \texttt{abstract}, \texttt{final}, or \texttt{default} modifiers, which indicate different override strategies. Subsequently, the \texttt{generateOverrides} function in Algorithm~\ref{algo:gen-override} ensures that the newly generated class overrides the \dquote{must override} and randomly overrides the \dquote{can override} methods. Algorithm~\ref{algo:gen-override} begins by constructing an inherited method map, where each key is the signature of an inherited method, and each value is a set of directly inherited methods along with their supertypes. If a newly generated IR class is an interface, we generate overridden methods randomly because both \dquote{override} and \dquote{not override} are legal. For a class, the rules in Table~\ref{tbl:override} are followed to determine whether to override a method. Finally, the overridden methods are added to the newly generated class.

\input{table-override}

Table~\ref{tbl:override} shows the universal override rules we designed for all four JVM-based languages to generate valid override methods in a subclass. The first two columns indicate whether the super method in the parent class is \texttt{Null}, \texttt{Abstract}, \texttt{Final}, or \texttt{Normal}. Algorithm~\ref{algo:gen-override} checks these method modifiers according to the column order: it first checks whether the super method is \texttt{Null}, then whether it is \texttt{Abstract}, and so on. The first row of this table indicates that when there are no super methods in the parent class and multiple methods (either abstract or concrete) with the same signature exist across multiple interfaces, the subclass \smallcapsbold{MUST} override the method with the same signature. If there are no super methods in the parent class and only one abstract method exists in the interfaces, the subclass \smallcapsbold{MUST} override that method. However, if the only method is concrete, the subclass \smallcapsbold{CAN} override the method. 

The second row presents cases where the parent class contains one abstract method. In such cases, the subclass \smallcapsbold{must} override the abstract method, regardless of whether there are concrete or abstract methods with the same signature in the interfaces. The third row describes cases where the parent class has one final method. Here, the subclass \smallcapsbold{cant} override the method if there are abstract methods or no methods with the same signature in one or more interfaces. If the method in the parent class is final and at least one interface method is concrete, the subclass cannot override the method. However, if the subclass’s language tag is Kotlin, the generated program is invalid (see Figure~\ref{fig:example2}), as Kotlin prohibits this scenario regardless of whether the method is overridden. In this scenario, if the subclass’s language tag is Kotlin, we adjust it to Java using the \textit{LangShuffler} mutator. By converting the Kotlin code to Java, the invalid Kotlin code can be avoided, making Kotlin align with other programming languages. We mark this case as \smallcapsbold{cant*} in Table~\ref{tbl:override}.

The last row presents cases where the parent class contains one normal method. The subclass \smallcapsbold{can} override the method if the interfaces contain only abstract methods or no methods with the same signature. If the interfaces contain any concrete methods, Java and Kotlin apply different override rules. In Java, the subclass can override the method because Java prioritizes methods in the class over those in interfaces. However, in Kotlin, class and interface methods are treated at the same level, so the subclass \smallcapsbold{must} override both concrete and abstract methods to resolve the conflict. When designing the overriding rules for all JVM languages, we adopted the most restrictive scenarios to ensure compatibility across all languages. We mark this case as \smallcapsbold{must} in Table~\ref{tbl:override}.

By default, each class has a 0.3 probability of having a parent class, and the number of interfaces that a class (or an interface) can implement (or extend) ranges from 0 to 3. We do not impose restrictions on the inheritance depth of classes and interfaces. Regarding the parameters and return value types of member methods, the \textit{Program Generator} tool will randomly select from the already generated classes or from predefined classes (e.g., the built-in class ``\texttt{kotlin.Any}''). Following the above process, the IR program generated by Algorithm~\ref{algo:overall} is valid and should pass the compilers as it follows the basic syntax rule of the four languages on JVM. We translate the IR program to different programming languages according to its language tag and obtain the cross-language program.

%% file: algo1.tex
\begin{algorithm}
\caption{Algorithm for generating cross-language IR program}
\label{algo:overall}
\SetKwProg{Fn}{fun}{\string:}{}
\SetKwFor{For}{for}{\string then}{}
\SetKwIF{If}{ElseIf}{Else}{if}{then}{elif}{else}{}
\SetKwFunction{FGenProgram}{generateIRProgram}
\SetKwFunction{FGenOverrides}{generateOverrides}
\Fn{\FGenProgram{}}{
$program \gets newProgram()$\;
        \For{$i = 1$ to $randomChosenNumberOfDeclarations$}{
            $class \gets generateClass()$\;
            
            
            $class.languageTag \gets randomLanguage()$\;
            $class.typeParameters \gets generateTypeParameters(class)$\;
            \tcc{Type parameters in current class can be used as type argument in super}
            $class.superTypes \gets generateSuperTypes(class)$\;
            $class.methods \gets generateMethods(class)$\;
            \tcc{See \FGenOverrides{class} in Algorithm~\ref{algo:gen-override}}
            $class.overrideMethods \gets$ \FGenOverrides{class}\;
            $program.addClass()$\;
        }
\KwRet{$program$} 
}
\BlankLine
\end{algorithm}
    

%% file: algo2.tex
\begin{algorithm}
\caption{Algorithm for generating override methods}
\label{algo:gen-override}
    \SetKwProg{Fn}{fun}{\string:}{}
    \SetKwFor{For}{for}{\string then}{}
    \SetKwIF{If}{ElseIf}{Else}{if}{then}{elif}{else}{}
    \SetKwFunction{FGenOverrides}{generateOverrides}
    \SetKwFunction{FCollectSignature}{collectMethodSignatureMap}
    \Fn{\FGenOverrides{class}}{
        $overrideMethods \gets newList()$\;
        \tcc{Collect a map where each key is the signature of an inherited method, and each value is a set of directly inherited methods along with their supertypes.}
        \For{$(signature, methods) \in collectMethodSignatureMap(class)$}{
            $(methodInSuperClass, methodsInInterface) \gets methods$\;            
            $overrideType \gets $ choose override type from Table~\ref{tbl:override}\;
            \If{$(overrideType == canOverride \land randomBoolean()) \lor overrideType == mustOverride$}{
                $method \gets generateMethodThatOverrides(methods)$\;
                $overrideMethods.add(method)$\;
            }
        }
        \KwRet{$overrideMethods$}\;
    }
\end{algorithm}

%% file: table-override.tex
\begin{table}
\newcommand{\Must}{\smallcapsbold{must}}
\newcommand{\Cant}{\smallcapsbold{Cant}}
\newcommand{\Can}{\smallcapsbold{Can}}
\newcommand{\Further}{\smallcapsbold{Further}}
\centering
\caption{The Universal Override Rules for Generating Valid Override Methods in a Subclass}
\vspace*{-3mm} 
\begin{tabular}{|m{0.08\linewidth}|>{\centering}m{0.08\linewidth}|>{\centering}m{0.14\linewidth}>{\centering}m{0.14\linewidth}|>{\centering}m{0.12\linewidth}>{\centering}m{0.12\linewidth}|>{\centering\arraybackslash}p{0.12\linewidth}|} 
\hline

\multicolumn{2}{|>{}m{0.18\linewidth}|}{\multirow{2}{0.3\linewidth}{}} & \multicolumn{2}{>{\centering}m{0.3\linewidth}|}{Multiple Methods in Multiple Interfaces} & \multicolumn{2}{>{\centering}m{0.24\linewidth}|}{One Method in One Interface} & \multirow{2}{\linewidth}{\centering No Method in Interfaces} \\ \cline{3-6}

\multicolumn{2}{|m{0.18\linewidth}|}{} & All methods are abstract & At least one method is concrete & The method is abstract & The method is concrete &  \\  \midrule
\multirow{4}{\linewidth}{\centering The super method in the parent class} 
  & Null    & \Must & \Must & \Must & \Can & --- \\ [2mm] 
 & Abstract & \Must & \Must & \Must & \Must & \Must \\[2mm] 
 & Final & \Cant & \Cant* & \Cant & \Cant* & \Cant \\[2mm] 
 & Normal & \Can & \Must & \Can & \Must & \Can \\[2mm] 
\hline
\end{tabular}
\label{tbl:override}
\end{table}

%% file: approach-test.tex
\subsection{Conduct Normal and Differential Test for Compilers}
\label{sec:approach-test}

The cross-language program generated in the previous step is considered valid because it adheres to the IR, which is defined according to the JVM language specifications. In cases of discrepancies, we downgrade the program to a more basic level to ensure compatibility. Therefore, it should pass the compiler check. In this phase, we conduct normal testing for the compilers by feeding the generated program to the latest compilers for Kotlin, Scala, Groovy, and Java, and checking whether they accept the program. Since the generation process does not involve constructing executable code, we cannot validate the functional correctness of the compiled program through execution. As a result, we do not verify the runtime behavior of these programs. If the program fails to pass the compiler check during this testing phase, it is flagged and passed on to the next analysis step for further evaluation.

Meanwhile, we design three light-weight mutators, which only involve syntax-level modifications, as shown in Figure~\ref{fig:framework}, to generate more diverse programs. These mutators are:

\begin{itemize} 
\item \textit{LangShuffler}: Change the IR language tag of one class. This mutation helps test how compilers handle the interaction between different programming languages.
\item \textit{FunctionRemoval}: Remove one function in a class or an interface. This mutation helps evaluate how compilers handle missing or incomplete function definitions, especially in overriding scenarios.
\item \textit{TypeChanger}: Change the type parameter. This mutation helps evaluate how compilers handle type-related changes, which are crucial in strongly-typed JVM-based languages.
\end{itemize}

As our framework targets cross-language program generation, including \textit{LangShuffler} is natural for testing the compiler’s handling of such scenarios. Given that the IR currently supports only classes, methods, and generic types, the mutation possibilities involve modifying methods, types, or inheritance structures. \textit{FunctionRemoval} targets method deletion, which can potentially break override relationships. We do not include a function addition mutator, as adding uniquely named methods rarely affects validity, while adding conflicting ones can clearly render programs invalid. \textit{TypeChanger} alters parameter, return, or generic types—we group all type changes under one mutator since our goal is to expose compiler bugs, not classify them. Another potential mutator could involve adding or removing a parent class. However, adding a parent class without changing methods is similar to method removal, while removing a parent class generally does not affect program validity. After one of the three mutations is applied, the validity of the mutated program cannot be guaranteed and the cross-language program becomes unverified, consequently, we cannot apply normal testing for the mutated program. Instead, we employ differential testing for the mutated program, which involves using two versions of independent compilers to evaluate the mutated program. The goal is for both versions of the compilers to agree on their compilation outcomes. Specifically, both compilers must either successfully compile the program, indicating that the mutation does not introduce any errors, or both must reject the program, indicating that the mutation causes an error that prevents successful compilation. In either case, this agreement ensures that both compilers are functioning the same. However, if one compiler successfully compiles the program while the other fails, this discrepancy indicates a potential issue with one of the compilers. In such cases, we know that the program triggers the bug in at least one compiler.

\begin{figure}
\centering
\begin{minipage}[c]{0.8\textwidth}
\begin{lstlisting}[language=Java]
// FILE: A0.java
public abstract class A0<T> {
    public abstract Object func(A0<Object> arg0, T arg1);
}
// FILE: A1.java
public abstract class A1 extends A0<A1> {
    public Object func(A0<A1> arg0, A1 arg1) {
        return null;
    }
}
// FILE: main.kt
class A2: A1()
\end{lstlisting}
\end{minipage}
\caption{\kt{74148}: One latest version of Kotlin compiler accepts this program, while another version does not}
\label{fig:kt74148}
\end{figure}

In most differential testing cases, when the results of two compilers differ, the well-defined semantics of the source code can help us easily identify the buggy compiler. For example, in Figure~\ref{fig:example2-1}, we can directly determine that an earlier version of the Kotlin compiler, which fails to compile this program, is buggy, while the latest version, which successfully compiles it, is correct. However, in a small number of challenging cases, further investigation and discussion are required to pinpoint which compiler contains the bug. As shown in Figure~\ref{fig:kt74148}, the code fails to compile in an earlier version of the Kotlin compiler but compiles successfully in the latest version. From the perspective of source code semantics, the function in \texttt{class A1} does not override the function in \texttt{class A0}. However, due to type erasure, identifying which compiler is buggy becomes more complex. In such cases, the program is flagged, passed to the minimization step, and forwarded to the development team for further investigation.

%% file: approach-manual.tex
\subsection{Identify Minimized Program to Trigger Compiler Bug}
When a program that may trigger a compiler bug is detected through normal testing or differential testing, it needs to undergo minimization. The purpose of this process is to reduce the complexity of the program, making it easier for compiler developers to pinpoint the root cause more efficiently and streamlining subsequent debugging and analysis workflows.

The minimization process is performed manually. Since compiler error messages often point to issues in specific methods (e.g., unimplemented methods or conflicting method definitions), we first prioritize removing unrelated methods during minimization and retain only the core methods that trigger the bug. Second, to reduce the complexity of cross-language inheritance (defined as the complexity increased by 1 for each switch in programming language along the inheritance chain from the base class to subclasses—e.g., Java parent → Kotlin child → Java grandchild, resulting in a total complexity of 2), we manually adjust the implementation language of classes. This involves rewriting a Kotlin subclass into Java code, for example, while preserving the original semantics, in order to minimize language transitions.

Additionally, during minimization, the following operations may be applied to function parameters and return types, in order to further eliminate complex type structures unrelated to the underlying issue.

\begin{itemize}
    \item Replace Custom Types: Substitute custom types (e.g., user-defined classes or interfaces) with language-agnostic types (e.g., String, Int);
    \item Simplify Generics:
    \begin{itemize}
        \item Replace generic type parameters (e.g., <T>) with concrete types (e.g., String);
        \item Remove generic parameters entirely to see if the program can still trigger the bug.
    \end{itemize}
\end{itemize}

During the minimization process, if removing or modifying a code snippet causes the original bug to no longer be triggered, the change is rolled back, and the focus shifts to minimizing other parts. If code reduction or language adjustments introduce new bugs (i.e., the compiler reports different error messages), the current program is marked as a new triggering case, and a new round of minimization is initiated. Once no further minimization steps can eliminate the bug, we label the current program as the minimized triggering program and submit it to the compiler developers.

%% file: experiment.tex
\section{Experiment Setup}
\label{sec:experiment}

\subsection{Compiler Selection} 
To avoid reporting previously known bugs, we target testing the latest development version of compilers. Note that our testing efforts were incremental, i.e., we concurrently developed \toolName{} and tested the compilers. Hence, we have run \toolName{} in its full capabilities for only one month. Table~\ref{tbl:cversions} shows the versions of the tested compilers. The latest version is used for normal testing, while the latest and earlier versions are used for differential testing. 
\input{table-compiler-versions}

\subsection{Fuzzing Process}
Currently, our fuzz testing does not involve an iterative process. This means there is no connection or dependency between the newly generated programs in each round and those from previous rounds. Therefore, the system does not need to preserve intermediate states during the testing process; it only needs to store the generated programs during the test reporting phase.

\subsection{Research Questions}
Our goal is to assess the robustness of compilers in handling cross-language compilation and the associated challenges. Our evaluation is based on the following research questions (RQs):

\textbf{RQ1. Is \toolName{} effective in finding bugs in JVM compilers?} The answer to this RQ determines whether \toolName{} can successfully identify meaningful compiler bugs, particularly those that may not be discovered through conventional testing methods.

\textbf{RQ2. What are the characteristics of the bug-revealing trigger programs?} The answer to this RQ provides deeper insights into the nature of the test cases that expose compilers' bugs. The identified characteristics of trigger programs can help developers understand compilers' weaknesses. 

\textbf{RQ3. What is the root cause of the found bugs?} This RQ addresses fundamental issues in JVM-based cross-language complication. By understanding the causes of cross-language compilation bugs, we aim to understand the responsibility of different compilers in this process.


\textbf{RQ4. How effective are the \textit{TypeChanger}, \textit{FunctionRemoval}, and \textit{LangShuffler} mutators, as well as the original program generator and minimization process, in detecting bugs?} This RQ evaluates the contribution of specific mutators and processes in \toolName{} to the overall bug detection capability.

%% file: table-compiler-versions.tex
\newcommand{\ktDev}{\href{https://github.com/JetBrains/kotlin/commit/9df698ee9f3693d2002b269ab944d0f5dcb75481}{dev-9df698e}}
\begin{table}[!ht]
    \centering
    \caption{Versions of Tested Compilers}
    \label{tbl:cversions}
    \begin{threeparttable}
    \begin{tabular}{l|p{4.5cm}|p{4.5cm}}
    \hline
        Compiler & Latest Version & Earlier Version \\ \hline
        Kotlin & \ktDev{} with K2~\tnote{a} & \ktDev{} with K1 \\ \hline
        Scala & 3.6.4-RC1-bin-20241231-1f0c576-NIGHTLY & 2.13.15 \\ \hline
        Groovy & 5.0.0-alpha-11 & 4.0.24 \\ \hline
        Java   & 17 & 1.8 \\ \hline
    \end{tabular}
    \begin{tablenotes}
      \item[a] K2 here means language version 2.0 and K1 means language version 1.9. Language version is a Kotlin compiler setting that enables different compiler implementations. Here language version 2.0 enables new compiler implementation and language version 1.9 enables the earlier implementation.
    \end{tablenotes}
    \end{threeparttable}
\end{table}

%% file: results.tex
\section{Results}
\label{sec:results}
\input{rq1}
\input{rq2}
\input{rq3}

\input{rq4}

%% file: rq1.tex
\subsection{Bug-Finding Results}
\label{rq1}
\input{tbl-results}

Once we located and minimized a program that produces different outputs in two versions of the compiler, we first searched for keywords in compilers' informative diagnostic messages through the bug tracking system to check for similar existing bugs and avoid reporting duplicates. After inspection, we reported the trigger program and the differing behaviors of the compilers to the development teams.

Table~\ref{tbl:results} lists the \countAll{} bugs found by \toolName{} in five JVM compilers: kotlinc, groovyc, scala3c, scala2c, and javac. The second column shows that all reported bugs in these compilers have been confirmed, with two of them being duplicates. Notably, most of the bugs in kotlinc were discovered during the \toolName{}'s implementation process. We applied an unfinished version of \toolName{} to test the compilers while simultaneously enhancing its ability to generate richer syntax features. The bugs in Groovy, Scala3, Scala2, and Java were discovered within a month after the tool stabilized. The number of detected bugs may not be large compared to other fuzzing approaches for two reasons. First, we specifically targeted inheritance and overridden typing bugs, which is a narrower scope. Second, as mentioned in Section~\ref{sec:approach-test}, our differential testing followed a strict rule: we only considered cases where one compiler successfully compiled the generated program while the other version failed to compile it. This way significantly reduced the number of detected bugs but also greatly minimized our manual inspection effort.

Another interesting observation from Table~\ref{tbl:results} is that, although both the Groovy and Kotlin development teams were very responsive to our bug reports, the Groovy developers addressed them more promptly. The Groovy team quickly fixed all four reported bugs, whereas the Kotlin team confirmed the bugs, but most of these bugs remain in the \dquote{confirmed} status and have not yet been fixed (only one has been fixed). We discussed this with the Kotlin team, and they explained that: \dquote{\textit{Kotlin targets multiple different platforms. On some of them, we can successfully implement and compile the test program into a perfectly valid executable. This means that the bug case could either be a JVM-specific issue or a general restriction of Kotlin, and we have not yet decided which approach is better.}} Since Groovy runs exclusively on the JVM, its development team only needs to focus on JVM-specific behavior. This may explain the different bug-fixing rates between Groovy team and Kotlin team.

The second-to-last column of Table~\ref{tbl:results} shows the severity of the reported bugs. The severity of bugs in Scala compilers is not available, so it is marked as \dquote{Not Specified} in the table. In groovyc, three out of four reported bugs are labeled \dquote{Major}. Additionally, in kotlinc, 6 out of 7 bugs are labeled \dquote{Normal}, and 1 is labeled \dquote{Major}. This demonstrates that the bugs found by \toolName{} are not trivial. Furthermore, the last column indicates the versions of the respective compilers affected by these bugs. While these bugs may also impact older versions, most compilers—except for javac—only maintain a few of the latest or most recent distributions. Therefore, only the latest or most recent affected versions are listed in Table~\ref{tbl:results}.


\begin{center}
\noindent\fbox{%
     \parbox{0.98\linewidth}{%
        \textbf{To answer RQ1, the bug-finding results demonstrate that \toolName{} can effectively detect compiler bugs in all five JVM-based compilers. Since these bugs also affect the latest versions, they deserve special attention.}
    }
    }%
\end{center}

%% file: tbl-results.tex
\newcommand{\defaultSvrt}{Not Specified}
\begin{table}[!ht]
    \centering
    \begin{threeparttable}
    \caption{Overview of Compiler Bugs found by \toolName{}}
    \label{tbl:results}
    \begin{tabular}{ccccc} \hline
Compiler  &  Bug ID     &  Status         &Severity & Affected Versions~\tnote{a}   \\ \hline
\multirow{11}{*}{kotlinc}
   &    \kt{74109}      &  Confirmed      & Normal  &    2.1.0               \\
   &    \kt{74147}      &  Fixed          & Major  & 2.1.20-Beta1, 2.0.0 \\
   &    \kt{74148}      &  Confirmed      & Normal &  2.1.20-Beta1, 2.0.0           \\
   &    \kt{74151}      &  Confirmed      &\defaultSvrt&  2.0.0      \\
   &    \kt{74156}      &  Duplicate      &\defaultSvrt&  2.1.20-Beta1, 1.9.25       \\
   &    \kt{74160}      &  Confirmed      &\defaultSvrt&  2.1.0, 1.9.25       \\
   &    \kt{74174}      &  Confirmed      &\defaultSvrt&  2.1.0          \\
   &    \kt{74188}      &  Confirmed      & Normal &  2.1.0          \\
   &    \kt{74202}      &  Now \jdk{8352290}~\tnote{b}& Normal & 2.1.0, 1.9.25\\
   &    \kt{74209}      &  Confirmed      & Normal &2.1.0\\
   &    \kt{74288}      &  Duplicate      & Normal &2.1.0, 1.9.25\\ \hline
\multirow{4}{*}{groovyc} 
   & \groovy{11548}     &  Fixed          & Major  &  4.0.24                  \\
   & \groovy{11549}     &  Fixed          & Minor  &  5.0.0-alpha-11, 4.0.24                \\
   & \groovy{11550}     &  Fixed          & Major  &  5.0.0-alpha-11, 4.0.24       \\
   & \groovy{11579}     &  Fixed          & Major  &  5.0.0-alpha-12, 4.0.26, 3.0.24 \\ \hline
\multirow{7}{*}{scala3c}   
   & \scalaThree{22307} &  Confirmed      &\defaultSvrt&  3.6.3-RC2     \\ 
   & \scalaThree{22308} &  Confirmed      &\defaultSvrt&  3.6.3-RC2, 3.6.4-RC1             \\ 
   & \scalaThree{22309} &  Confirmed      &\defaultSvrt&  3.6.3-RC2, 3.6.4-RC1              \\ 
   & \scalaThree{22310} &  Confirmed      &\defaultSvrt&  3.6.3-RC2, 3.6.4-RC1           \\ 
   & \scalaThree{22311} &  Confirmed      &\defaultSvrt&  3.6.3-RC2, 3.6.4-RC1          \\ 
   & \scalaThree{22312} &  Confirmed      &\defaultSvrt&  3.6.3-RC2, 3.6.4-RC1                \\ 
   & \scalaThree{22717} &  Confirmed      &\defaultSvrt&  3.6.3-RC2               \\ \hline
\multirow{2}{*}{scala2c}
  & \scalaTwo{13074}    &  Confirmed      &\defaultSvrt& 2.13.15            \\ 
  & \scalaTwo{13075}    &  Confirmed      &\defaultSvrt& 2.13.15             \\ \hline
\multirow{2}{*}{javac}      
& \jdk{8347330} & Not an issue~\tnote{c}  & - & -          \\
& \jdk{8352290} & Confirmed & Minor & 8\\
\hline
        
    \end{tabular}
    \begin{tablenotes}
      \item[a] Except for javac, other compilers only maintain several the latest or most recent distributions, so only the latest or most recent versions affected are listed here.
      \item[b] This issue was first submitted to the Kotlin team and assigned an issue ID, but it was later confirmed as a javac bug after extensive tracking and verification. It has been documented in \jdk{8352290}. We will further discuss this bug in Section~\ref{rq3}.
      \item[c]Although the Java team does not currently consider this issue to be a bug, this example still has some impact (Indirectly caused \kt{74148}, \kt{74174} and \scalaTwo{13074}).
    \end{tablenotes}
\end{threeparttable}
\end{table}


        

%% file: rq2.tex
\subsection{Trigger Program Characteristics}
\label{rq2}

Before minimization, all trigger programs were cross-language, involving at least one language referencing another. After manual minimization, some trigger programs became single-language, indicating that cross-language interactions are not the root cause of those bugs. For trigger programs that remained cross-language after minimization, we further reduced irrelevant code parts during the rollback operation. Table~\ref{tbl:trigger-prog} shows the characteristics of trigger programs.

\input{table-trigger-program}

\subsubsection{Single Language or Cross-Language in Trigger Programs} 
\label{sec:rq2-single-or-cross}
The first two rows of Table~\ref{tbl:trigger-prog} present whether the trigger programs are cross-language or single-language. It shows that 7 trigger programs are single-language, while the remaining 17 out of \countAll{} are cross-language.

Figure~\ref{fig:single-lang-1} presents a Kotlin-only program that triggers a bug in the latest Kotlin compiler. In this case, \texttt{class A2} implements \texttt{interface I1} with different type parameters, \texttt{I0} and \texttt{A2}. The latest Kotlin compiler incorrectly accepted this program, whereas earlier versions correctly reported the error \dquote{INCONSISTENT_TYPE_PARAMETER}. This indicates a regression in the compiler's type-checking mechanism and the Kotlin team has since fixed this bug. 

Figure~\ref{fig:single-lang-2} illustrates another single-language bug in the Scala3 compiler. In this case, the \texttt{func} method in \texttt{class A1} has a compile-time signature that differs from \texttt{func} in \texttt{class A0}. As a result, \texttt{A1::func} is not actually an override of \texttt{A0::func}. However, the Scala3 compiler incorrectly accepted this invalid program, whereas the Scala2 compiler reported an \dquote{override nothing} error correctly. This discrepancy highlights a potential regression or inconsistency in the Scala3 compiler’s handling of method overrides.

\begin{figure*}[t!]
\centering
\begin{subfigure}[t]{0.35\textwidth}
\begin{lstlisting}[language=Kotlin]
// All Kotlin Classes 
interface I0
interface I1<T0>: I0
interface I2: I1<I0>
class A2: I2, I0, I1<A2>
\end{lstlisting}
\caption{}
\label{fig:single-lang-1}
\end{subfigure}%
~
\begin{subfigure}[t]{0.55\textwidth}
\begin{lstlisting}[language=Scala]
// All Scala Classes
abstract class A0[T] {
  def func(arg0: A0[String], arg1: T): Unit
}
abstract class A1 extends A0[String] {
  override def func(arg0: A0[Object], arg1: String): Unit = {}
}
\end{lstlisting}
\vspace*{-2mm} 
\caption{}
\label{fig:single-lang-2}
\end{subfigure}
\caption{(a)\kt{74147}: Kotlin compiler accepts an invalid program with inconsistent type parameter values; \\(b)\scalaThree{22310}: Scala3 compiler accepts an invalid program when a method overrides nothing}
\end{figure*}



Figure~\ref{fig:multi-lang-1-1} shows a Java-Groovy program that triggers a bug in the Groovy compiler. The method \texttt{func} in \texttt{class B} overrides \texttt{func} in class \texttt{A}. However, the Groovy compiler rejects this valid program and reports the error: \dquote{Can't have an abstract method in a non-abstract class. The class \squote{C} must be declared abstract or the method \squote{java.lang.Object func()} must be implemented}. The Groovy compiler wrongly interpreted that \texttt{B.func} is not an override. Furthermore, during minimization, we found that removing the type parameter in \texttt{class A} (renaming it to \texttt{A1} for distinction, as shown in Figure~\ref{fig:multi-lang-1-2}) causes the bug to disappear. This is because, on the JVM, the method \texttt{B.func} semantically overrides \texttt{A.func}. However, in the bytecode, the actual override occurs through a bridge method, which has the same signature as \texttt{A.func} after type erasure. Since \texttt{class B1} lacks a bridge method due to the absence of a type parameter, we inferred that the bug might be related to bridge method generation and reported it to the Groovy developers. The \href{https://github.com/apache/groovy/commit/e484ac5a3e85b25bb64247b84f56a047606cc54b}{6cc54b} fix for this bug confirms our assumption that the issue was indeed related to the bridge method. This demonstrates that not only can \toolName{} generate programs that trigger compiler bugs, but it can also help identify the root cause and even suggest a focal fixing point through the minimization process.



\begin{figure*}[t!]
\centering
\begin{subfigure}[t]{0.45\textwidth}
\begin{lstlisting}[language=Java]
// Java Classes
public interface A<T> {
    public T func();
}
public interface B extends A<String> {
    @Override
    default String func() {
        return "I1";
    }
}
// Groovy Class
public class C implements A<String>, B {}
\end{lstlisting}
\vspace*{-2mm} 
\caption{}
\label{fig:multi-lang-1-1}
\end{subfigure}%
~ 
\hspace{2em}
\begin{subfigure}[t]{0.38\textwidth}
\begin{lstlisting}[language=Java]
// Java Classes
public interface A1 {
    public String func();
}
public interface B1 extends A1 {
    @Override
    default String func() {
        return "I1";
    }
}
// Groovy Class
public class C1 implements A1, B1 {}
\end{lstlisting}
\vspace*{-2mm} 
\caption{}
\label{fig:multi-lang-1-2}
\end{subfigure}
\caption{(a)\groovy{11549}: Groovy compiler rejects a valid program when implementing Java interfaces;\\(b) Remove the type parameter in Line 5 and 12, the bug disappears.}
\end{figure*}

Figure~\ref{fig:rq2-multi1} shows a Kotlin-Java program that triggers a bug in the Kotlin compiler. All five Java classes compile successfully using the Java compiler. However, when the Kotlin \texttt{class Child} extends \texttt{Parent} and implements \texttt{IChild}, the Kotlin compiler rejects the program, reporting: \dquote{Class \squote{Child} must override \squote{func} because it inherits multiple implementations for it}. During minimization, if we translate the Kotlin \texttt{class Child} into Java, the program compiles correctly. This indicates that the bug lies in the Kotlin compiler, which incorrectly handles the Java code.


\begin{figure}
\begin{minipage}[c]{0.8\textwidth}
\begin{lstlisting}[language=Java]
// Java Classes
public interface ITop {
    public default void func() {}
}
public interface ISecondary extends ITop {
    public default void func() {}
}
public interface IChild extends ISecondary, ITop {}
public class GrandParent implements ITop {
    public final void func() {}
}
public class Parent extends GrandParent implements ISecondary {}
// Kotlin Class
abstract class Child : Parent(), IChild
\end{lstlisting}
\end{minipage}
\caption{\kt{74109}: Kotlin compiler rejects a valid program when a Kotlin class extends a Java class}
\label{fig:rq2-multi1}
\end{figure}

\subsubsection{Generics Characteristics in Trigger Programs} 
\label{sec:rq2-generics}
The third and fourth rows of Table~\ref{tbl:trigger-prog} indicate whether the trigger programs are related to generics. As shown, 20 out of \countAll{} trigger programs involve generics and contain type parameters, highlighting that a significant portion of the discovered bugs are associated with type parameter interactions. This suggests that generic types play a crucial role in exposing compiler inconsistencies, particularly in cross-language scenarios. Examining the code examples in Section~\ref{sec:rq2-single-or-cross}, Figures~\ref{fig:single-lang-1}, \ref{fig:single-lang-2}, and \ref{fig:multi-lang-1-1} all include type parameters, while Figure~\ref{fig:rq2-multi1} is the only one without type parameters. This observation aligns with recent research suggesting that well-typed programs can still exhibit errors~\cite{chaliasos2021well,chaliasos2022finding}. However, while prior studies primarily focus on a single JVM compiler, our study results extend this insight by revealing that interactions between well-typed languages in cross-language scenarios can also lead to unexpected compiler failures.


\subsubsection{Inheritance Characteristics in Trigger Programs} The last two rows of Table~\ref{tbl:trigger-prog} indicate the average inheritance depth and width of the trigger programs. The inheritance depth is defined as the maximum length of an inheritance chain in a trigger program, while the inheritance width refers to the maximum number of classes or interfaces in a single class or interface signature. For example, the inheritance depth and width of the trigger program in Figure~\ref{fig:single-lang-1} are 3 and 3, respectively, while in Figure~\ref{fig:multi-lang-1-1}, they are 2 and 2, respectively. 

As shown in Table~\ref{tbl:trigger-prog}, the average inheritance depth of trigger programs for kotlinc, scala3, and scala2 exceeds 3.29, and the average inheritance width for kotlinc, groovyc, and scala2 is greater than 1.5. This suggests that triggering compiler bugs in cross-language scenarios often requires relatively complex programs with deeper inheritance hierarchies or wider class/interface signatures. These two metrics provide valuable insights into the compilers' ability to handle override scenarios and manage intricate type relationships across different JVM-based languages.

\begin{center}
\noindent\fbox{%
     \parbox{0.98\linewidth}{%
        \textbf{To answer RQ2, while 7 of the trigger programs that reveal compiler bugs contain code in a single language, 17 out of \countAll{} contain cross-language code. These programs involve type parameters and complex inheritance structures, demonstrating their effectiveness in testing compilers' ability to handle cross-language compilation.}
    }
    }%
\end{center}


%% file: table-trigger-program.tex
\begin{table}
\centering
\caption{The Characteristics of Trigger Programs}
\begin{tabular}{lcccccc} \hline
 & kotlinc & groovyc   & scala3 & scala2 & javac & Total \\ 
\hline
Cross-Language & 8 & 1 & 5 & 2 & 1 & 17 \\
Single-Language & 2 & 3 & 2 & 0 & 0 & 7 \\ 
\hline
Generic Related & 8 & 3 & 6 & 2 & 1 & 20 \\
Not Generics Related & 2 & 1 & 1 & 0 & 0 & 4 \\ 
\hline
Average Inheritance Depth & 3.40 & 2.50 & 3.29 & 3.5 & 3 & 3.21\\
Average Inheritance Width & 1.90 & 1.75 & 1.43 & 1.5 & 1 & 1.67 \\
\hline
\end{tabular}
\label{tbl:trigger-prog}
\end{table}

%% file: rq3.tex
\subsection{Root Causes of the Found Bugs}
\label{rq3}

As mentioned in Section~\ref{rq2}, 7 out of \countAll{} trigger programs are written in a single language. This indicates that the corresponding bugs exist within a single compiler. We inspected these 7 bugs (\kt{74147}, \kt{74156}, \groovy{11548}, \groovy{11550}, \groovy{11579}, \scalaThree{22309}, \scalaThree{22310}) and found that 6 bugs are related to the compiler's semantic check module for inheritance or overriding conflicts, while \scalaThree{22309} is associated with the backend's code generation. These bugs exhibit failures in detecting method overriding/inheritance conflicts or anomalies in bytecode generation, which align with bugs identified by existing single-language fuzzing techniques~\cite{chaliasos2021well}.

Except for single-language programs, 17 out of \countAll{} trigger programs are cross-language, requiring two compilers to work together to compile the programs. In these scenarios, when the compilation result does not meet expectations, it is crucial to understand the intrinsic nature of these bugs and determine which compiler should be responsible for fixing them.

\begin{figure}[!th]
\centering
\begin{minipage}[c]{0.60\textwidth}
\begin{lstlisting}[language=Kotlin]
// FILE: I0.kt
interface I0<T0, T1> {
    fun func(t1: T0)
    fun func1(): T1
}
// FILE: A0.kt
open class A0<T2>: I0<T2, Nothing> {
    override fun func(t1: T2) {}
    override fun func1(): Nothing = throw UnsupportedOperationException()
}
// FILE: A1.java
class A1 extends A0<String> {}
\end{lstlisting}
\end{minipage}
\caption{\jdk{8352290} (Once was \kt{74202}): Java code cannot be compiled when inheriting from a Kotlin class with \texttt{Nothing} type parameter}
\label{fig:rq3-javac-bug}
\end{figure}

Figure~\ref{fig:rq3-javac-bug} shows a trigger program where the Java \texttt{class A1} extends the Kotlin \texttt{class A0} and implements the Kotlin interface \texttt{I0}. The \texttt{Nothing} type, which is unique to Kotlin's type system, is used as a type parameter in the supertype \texttt{I0<T2, Nothing>} of \texttt{class A0}. \texttt{Nothing} represents a value that never exists or a function that never returns (e.g., Line 9 in Figure~\ref{fig:rq3-javac-bug}). As a bottom type, \texttt{Nothing} is a subtype of all other types in Kotlin. This property allows it to be used in variance annotations (e.g., List<out T>) to safely handle type relationships. However, since Java lacks a similar concept, the Kotlin compiler cannot generate a reasonable generic representation for \texttt{Nothing}. So the supertype \texttt{I0<T2, Nothing>} of \texttt{class A0} provided to Java compiler is actually a raw type \texttt{I0}. The bytecode difference of \texttt{class A0} between code with and without \texttt{Nothing} is that \texttt{A0} either implements raw type \texttt{I0} or \texttt{I0<T2, ...>}. The Kotlin compiler can generate correct overrides for \texttt{I0::func} (either directly or as a synthetic bridge). In Java 1.8, the compiler does not think that \texttt{A0::func(T2)} overrides \texttt{I0::func(T0)} and wrongly reported \dquote{name clash: func(T2) in A0 and func(T0) in I0 have the same erasure, yet neither overrides the other}. This trigger program was initially reported to the Kotlin development team as \kt{74202}, as we believed that the issue was caused by the Kotlin compiler's inability to handle the lack of a proper representation for the \texttt{Nothing} type in Java. However, the Kotlin team labeled it as \dquote{Third Party Problem}. Subsequently, we reported the issue to the Java team, who confirmed it as a \jdk{8352290} bug in the \texttt{javac} compiler.

Another example in Figure~\ref{fig:kt74148} shows a Kotlin class inherits an abstract Java class (which inherits another Java class) with generics. In the cross-language compilation process, the rationale that javac not reporting an error is that \texttt{class A1} is not further extended. If there is a Java \texttt{class A3} extends \texttt{A1}, javac reports: \dquote{Class \squote{\texttt{A3}} must either be declared abstract or implement abstract method \squote{\texttt{func(A0<Object>, T)}} in \squote{\texttt{A0}} }. The key question is whether the error should be reported on the Kotlin side when the Kotlin \texttt{class A2} is implemented, or whether the Java compiler should be more proactive in enforcing the inheritance constraints between \texttt{class A0} and \texttt{A1}. Ultimately, the Kotlin team confirmed this as a Kotlin bug (\kt{74148}).

The above examples motivated our discussion regarding the root cause and responsibilities of different language compilers in addressing cross-language compilation bugs. Although the four JVM languages achieve good interoperability, their type systems and inheritance designs differ significantly. Java generally has looser inheritance restrictions and fewer defined types, while Kotlin enforces stricter inheritance rules and introduces additional types. When other JVM languages interact with Java, it is important for their compilers to align with each other. 

Furthermore, the two languages can be categorized as the provider and the user in a \textit{simple} cross-language scenario. For instance, if \texttt{class A} extends \texttt{class B}, \texttt{A} is the user and \texttt{B} is the provider. The challenges in cross-language compilation include, but are not limited to: 1. As a provider, how to translate new definitions into existing definitions on the Java side (e.g., \texttt{Unit} and \texttt{Nothing} in Kotlin and Scala)? 2. As a user, how to handle looser restrictions on the Java side? The ideal and general rule is: \dquote{User A has the responsibility to correctly use what is promised by B, and Provider B has the responsibility to correctly provide what is promised by its specification.} We investigate which compiler takes ownership of cross-language compilation bugs. After manually inspecting all 17 cross-language programs, we found that 9 bugs were confirmed on the user side and 5 on the provider side. Additionally, 3 bugs (\kt{74188}, \kt{74209}, \scalaThree{22312}) involve complex situations that cannot be easily classified as solely user- or provider-side issues. This observation aligns with the examples discussed in this section, reinforcing that, in most cases, the user side bears greater responsibility than the provider side.

\begin{center}
\noindent\fbox{%
     \parbox{0.98\linewidth}{%
        \textbf{To answer RQ3, most trigger programs in a single language reveal bugs in the compiler's semantic check module, particularly related to inheritance or overriding conflicts. On the other hand, cross-language trigger programs highlight that the user side (the language referencing a class/interface from another language) typically bears more responsibility to adapt to the provider's type system and inheritance rules. }
    }
    }%
\end{center}

%% file: rq4.tex
\subsection{Effectiveness of Mutators and Processes}
\label{rq4}

To evaluate the effectiveness of the mutators or the processes in \toolName{} that triggered the bug, we performed a detailed manual inspection of the triggering program and carefully examined the logs from each process or mutator involved. If the log indicates that the program does not trigger the bug before a particular process or mutator, but starts triggering the bug after the process or mutator has been executed, we conclude that the process or mutator is effective in triggering the bug. 

Figure~\ref{fig:rq4-langshuffler} illustrates a bug discovered by the \textit{LangShuffler} mutator. According to Table~\ref{tbl:override}, the method \texttt{A::func} is final, and the method \texttt{I0::func} is concrete; thus, the overriding rule for the method \texttt{func} in \texttt{class B} is \smallcapsbold{CANT*}. As introduced in Section~\ref{subsec:gen-program-alog}, the language tag of \texttt{class B} was set to Java. However, after applying the \textit{LangShuffler} mutator, \texttt{class B} in Java was changed to \texttt{class B} in Groovy, and this program revealed an inconsistency between two versions of the Groovy compiler, successfully triggering a bug in the Groovy compiler. Surprisingly, after minimization, we found the trigger program to be very simple. In fact, the structure of these three classes in \groovy{11548} is the same as in \kt{10195}, which we introduced in Figure~\ref{fig:example2} in Section~\ref{subsec:motivation-2}.

\begin{figure}
\begin{minipage}[c]{0.5\textwidth}
\begin{lstlisting}[language=Java]
// Groovy Classes
class A {
    final public void func() {}
}
interface I0 {
    public default void func() {}
}
class B extends A implements I0 {}
\end{lstlisting}
\end{minipage}
\vspace*{-2mm} 
\caption{\groovy{11548} (similar to \kt{10195} in Figure~\ref{fig:example2}): Groovy compiler rejects a valid program when an interface and parent class contain a method with the same signature}
\label{fig:rq4-langshuffler}
\end{figure}

Figure~\ref{fig:rq4-typechanger} illustrates a bug discovered by the \textit{TypeChanger} mutator. In the Scala \texttt{class A0}, the type parameter (Line 14) and method parameter (Line 15) were originally \texttt{String}. After applying the \textit{TypeChanger} mutator, the program revealed an inconsistency between two versions of the Scala 3 compiler, successfully triggering a bug in the Scala 3 compiler. Specifically, the mutated program is invalid because the mutated \texttt{class A0} exhibits conflicting base types: \texttt{I0[String]} and \texttt{I0[Object]}.

\begin{figure}
\begin{minipage}[c]{0.95\textwidth}
\begin{lstlisting}[language=Scala]
// Java Classes
public interface I0<T> {
    public default T func(T t) {
        return t;
    }
}
public interface I1 extends I0<String> {
    @Override
    default String func(String s) {
        return s;
    }
}
// Scala Class
abstract class A0 extends I1 with I0[Object /*String before mutate*/] {
  override def func(s: Object/*String before mutate*/): String = super.func(s)
}
\end{lstlisting}
\end{minipage}
\caption{\scalaThree{22311}: Scala3 compiler accepts an invalid program when extending a Java class}
\label{fig:rq4-typechanger}
\end{figure}

Figure~\ref{fig:example2-1} illustrates a bug discovered by the \textit{FunctionRemoval} mutator. Before this mutation, \texttt{class A3} contained an overridden \texttt{func} method, as the method \texttt{A1::func} in the parent class is a normal method and the interface method \texttt{I1::func} is concrete. Both versions of the Kotlin compiler passed this program. After the \textit{FunctionRemoval} mutator was applied and the overridden method \texttt{A3::func} was removed, one version of the Kotlin compiler passed the program while the other rejected it, triggering this bug.

Table~\ref{tbl:mutator} presents the statistics of mutators and processes in triggering bugs. Overall, differential testing discovered 22 out of \countAll{} bugs and is much more effective than normal test, which shows the effectiveness of mutators. Among all mutators, \textit{TypeChanger} is the most effective, identifying 11 out of \countAll{} bugs. This is understandable because different programming languages handle types differently, which can easily lead to inconsistencies or mismatches in cross-language interactions. The second most effective mutator is \textit{LangShuffler}, which discovered 5 out of \countAll{} bugs. By shuffling language, \textit{LangShuffler} can create programs that are syntactically valid but semantically incorrect. This can reveal bugs in the compiler's parsing, type-checking, or code generation phases. \textit{FunctionRemoval} ranks third, revealing 3 bugs. Additionally, the original program generator was able to uncover 2 bugs. These results demonstrate the effectiveness of the designed mutators and program generators.

We also observed that during the minimization process, 3 new bugs were identified. In addition to the three mutators used in the automatic process, we discovered another creative mutator during the manual minimization process: shuffling the order of interfaces implemented by a class. For instance, in Figure~\ref{fig:example2-1}, we found that changing the order of supertypes of \textbf{A3} in Line 15 to \dquote{class A3: I1, I0, A2<Any>()} caused the program to transition from being rejected to being accepted, even though this change should not alter the program's behavior. We reported this finding to the Kotlin developers, assisting them in locating and addressing the bug. This demonstrates that manual minimization can inspire creativity and is effective in uncovering bugs. We also plan to implement this mutator in our future work. 

\input{tbl-mutator}

\begin{center}
\noindent\fbox{%
     \parbox{0.98\linewidth}{%
        \textbf{To answer RQ4, \textit{TypeChanger} proves to be the most effective mutator in discovering cross-language compiler bugs, followed by \textit{LangShuffler} and \textit{FunctionRemoval}. Additionally, the original program generator and the creative manual minimization process also demonstrate effectiveness in finding compiler bugs.}
    }
    }%
\end{center}

%% file: tbl-mutator.tex
\begin{table}
\centering
\caption{Bugs Discovered by Each Mutator and Process}
\begin{tabular}{cccccccc} \hline
Test Strategy & Mutator/Process & kotlinc & groovyc   & scala3 & scala2 & javac & Total \\ 
\hline
\makecell{Normal \\ Testing}&\makecell{Original \\ Program Generator} 
                       & 0  & 2 & 0 & 0 & 0 & 2  \\
\hline
\multirow{4}{*}{\makecell{Differential \\ Testing}}
     &LangShuffler     & 1  & 1 & 3 & 0 & 0 & 5  \\ 
     &TypeChanger      & 5  & 1 & 3 & 1 & 1 & 11 \\
     &FunctionRemoval  & 3  & 0 & 0 & 0 & 0 & 3  \\
&Minimization  Process & 1  & 0 & 1 & 1 & 0 & 3  \\ 
\hline
Total &                & 10 & 4 & 7 & 2 & 1 & 24 \\
\hline
\end{tabular}
\label{tbl:mutator}
\end{table}

%% file: discussion.tex
\section{Discussion}
\label{sec:discussion}


\textbf{\textit{Analysis of Results}}: Unlike methods that select and mutate seeds from test corpora, \toolName{} generates programs entirely from scratch. To ensure that the generated programs do not deviate significantly from normal test code, structured IR and strict inheritance rules are enforced. Additionally, three effective mutators are designed to enhance cross-language program diversity. The results of \textbf{RQ1} and \textbf{RQ2} demonstrate that programs with diverse type parameters and complex inheritance structures are effective at triggering compiler bugs. Notably, 17 out of \countAll{} programs (70.8\% of all tested cases) involve cross-language compilation and successfully exposed compiler bugs in cross-compilation scenarios. This highlights the challenges that JVM-based compilers face when handling cross-language interoperability. The results of \textbf{RQ4} show that differential testing is more effective in triggering compiler bugs compared to normal test. While all processes and mutations contribute to bug discovery, the \textit{TypeChanger} mutator is the most effective in exposing compiler issues. One reason for this is the inherent differences in type systems across languages, making type alignment particularly challenging. \textbf{RQ3} further explores the responsibilities of the two compilers in cross-compilation bugs. The cross-language bug data points suggest that the user side — the language referencing an entity from another language (the provider) — tends to bear more responsibility for adapting to the provider’s type system.



\noindent
\textbf{\textit{Threats to Validity}}: The first threat to validity is that three new bugs were discovered during the manual minimization process, which requires significant domain knowledge and creativity. We cannot guarantee that these results can be reproduced. However, 21 out of \countAll{} (87.5\%) bugs can be identified through other automated processes, demonstrating the effectiveness of the \toolName{} framework. Another threat is that, before reporting a bug, we need to search issue tracking systems and manually check whether the program is similar to previous trigger programs to avoid duplicate reports. The persistent recurrence of already-reported bugs significantly increases our manual validation overhead. In this \dquote{needle-in-a-haystack} process, we may miss some low-reproducibility bugs due to the large number of recurring ones. We plan to automate this process in the next step. 



%% file: related.tex
\section{Related Work}
\label{sec:related}

\subsection{Compiler Fuzzing}
There is a plethora of work on constructing diverse test programs to evaluate compiler validity~\cite{chen2020survey}. The three major approaches are discussed in this section.


\subsubsection{Generated-based Compiler Fuzzing}
\texttt{Csmith} creates programs covering a large subset of C while avoiding undefined and unspecified behaviors that could hinder its ability to automatically detect incorrect code bugs \cite{csmith}. YARPGen proposed a method for generating expressive programs that avoid undefined behavior without relying on dynamic checks, as well as generation policies that enhance the diversity of generated code and trigger more optimizations~\cite{YARPGen}. YARPGen uncovered over 220 bugs in GCC, LLVM, and the Intel® C++ Compiler. Dewey \textit{et al}. have introduced a constraint logic programming approach for synthesizing Rust programs and identified 18 bugs in the Rust typechecker~\cite{dewey2014language,dewey2015fuzzing}. Hephaestus constructed programs more likely to trigger typing bugs and introduced two novel transformation-based approaches for uncovering type inference and soundness compiler bugs in Java, Kotlin, and Groovy compilers~\cite{chaliasos2022finding}. Georgescu \textit{et al}. generated random code snippets by leveraging Kotlin semantic and syntactic models, and introduced two genetic algorithms to diversify the testing program~\cite{georgescu2024evolutionary}. They conducted differential testing for different versions of Kotlin compiler and found bugs within the Kotlin compilers. Zang \textit{et al}. proposed a framework which allows developers to incorporate their domain knowledge on testing compilers, giving a basic program structure that allows for exploring complex programs that can trigger Java JIT compilers bugs~\cite{zang2022compiler,zang2024java}.

\subsubsection{Mutation-based Compiler Fuzzing}
Instead of generating complete programs from scratch, the main idea of program mutation is to modify parts of an existing test program. Equivalence Modulo Inputs (EMI) can achieve program mutation by stochastically pruning, inserting, or modifying its unexecuted code on the given input. These EMI variants can help differentially test GCC and LLVM compilers~\cite{le2014compiler}. Lidbury \textit{et al}. generated programs suitable for OpenCL compilers and inserted dead-by-construction code at random locations of the original program to perform semantics-preserving code mutations ~\cite{lidbury2015many}. Chen \textit{et al}. mutated class files with a wide range of mutation operations, such as inserting/deleting methods into/from classes, and performed a sampling to select mutations that have larger possibilities to trigger JVM bugs~\cite{chen2016coverage}. \texttt{JavaTailor} synthesizes diverse test programs by weaving elements extracted from historical bug-revealing JVM test programs into seed programs to cover more JVM behaviors and paths~\cite{zhao2022history}. \texttt{SJFuzz} schedules seeds (class files) for mutations based on the discrepancy and mutates class files via control flow mutators for diversifying class file generation to find JVM bugs~\cite{wu2023sjfuzz}.

\subsubsection{Learning-based Compiler Fuzzing}
\texttt{TreeFuzz} parses the training program into a tree structure and learns a probabilistic model that synthesizes new fuzzing inputs through tree traversal~\cite{patra2016learning}. \texttt{TitanFuzz} leverages Large Language Models (LLMs) to produce the initial seed programs for fuzzing and adopts an evolutionary strategy to produce new test programs by using LLMs to automatically mutate the seed programs for fuzzing DL libraries~\cite{deng2023large}. \texttt{WhiteFox} first uses an LLM-based analysis agent to examine the low-level optimization source code and produce requirements on the high-level test programs that can trigger the optimization; then uses an LLM-based generation agent to produce test programs based on the summarized requirements~\cite{yang2024whitefox}. \texttt{FUZZ4ALL} automatically distills user inputs into a prompt and then iteratively updates the initial input prompt with both code examples and generation strategies to obtain diverse and realistic inputs for any practically relevant language~\cite{xia2024fuzz4all}. \texttt{MetaMut} integrates the compiler domain knowledge into prompts and guides an LLM to generate natural-language mutator. Then it asks the LLM to fill a carefully crafted mutator template for synthesizing test programs for GCC and Clang compilers~\cite{ou2024mutators}. 

The above approaches have demonstrated notable effectiveness in fuzzing compilers with single-language programs. In contrast, we designed a universal IR for JVM-based languages and proposed a \toolName{} framework to generate multi-language programs, specifically targeting the testing of how compilers handle interactions among different JVM languages. We combined the generated-based and mutation-based approach and proposed a novel method for aligning overriding rules among different languages and introduced a new mutator, \texttt{LangShuffler}, to generate various interactions between different languages. Although a recent framework, \texttt{Atlas}, includes an automatic cross-language generator and can fuzz Android closed-source native libraries \cite{xiong2024atlas}, our \toolName{} framework is the first prototype specifically designed to test compilers' ability to handle compilations in JVM-based cross-language interactions.

\subsection{Cross-Language Defects}
Recent studies show that bugs in multilingual systems tend to propagate through the heterogeneous programming languages the systems are developed in~\cite{yang2023demystifying,yang2024learning}. Also the complexity of code changes for such bug fixes is significantly higher than that of fixes in single-programming-language scenarios~\cite{li2022vulnerability,li2023understanding}. Li \textit{et al}. found that in multi-language systems developers need to address bugs occurring in the interface of two languages, resulting in a higher number of lines of code~\cite{li2023understanding}. Abidi \textit{et al}. conducted an empirical study of 98 releases of nine open source JNI projects and concluded that files with multi-language design smells can often be more associated with bugs than files without these smells, and that specific smells are more correlated to fault-proneness than others~\cite{abidi2021multi}. \texttt{POLYCRUISE} uses holistic dynamic information flow analysis for multilingual software and discovered 14 unknown cross-language security vulnerabilities in real-world multilingual systems~\cite{li2022polycruise}. Based on code token similarity, structural similarity, and behavioral similarity, \texttt{COSAL} uses both static and dynamic analysis to support cross-language code-to-code search~\cite{mathew2021cross}. Youn \textit{et al}. extended \texttt{CodeQL} and developed a static analyzer to track data flows across Java-C and Python-C language boundaries and detect interoperation bugs for multilingual programs~\cite{youn2023declarative}. Li \textit{et al}. proposed 7 inter-language design smells for multi-language deep learning frameworks written in Python and C/C++, and extracted the detection rules for each design smell between Python and C/C++~\cite{li2025automated}. Feng \textit{et al}. conducted an empirical study for Kotlin-Java open source systems and revealed seven common code mistakes between Java and Kotlin~\cite{feng2024cross,feng2024depends}.

Beyond code issues in heterogeneous programming languages, cross-language compilation has also recently gathered increasing attention. Baradaran \textit{et al}. explored the challenges of cross-compiling a high-level language code base into WebAssembly~\cite{baradaran2024reusing}. Our work complements this by examining cross-language compilation among JVM-based languages. Our results reveal that, despite Java, Kotlin, and other JVM languages being well-typed, the JVM cross-language compilation process still encounters type-related bugs. The root cause lies in discrepancies of type systems and inheritance rules across different languages, making cross-language compilation error-prone.

%% file: conclusion.tex
\section{Conclusions and Future Work}
\label{sec:conclusion}

This paper introduces the \toolName{} framework, which generates cross-language programs using a universal JVM-based language intermediate representation (IR) and enhances program diversity through three specialized mutators aimed at exposing bugs in cross-language compilation. The generated and mutated programs incorporate various type parameters and complex inheritance structures, demonstrating their effectiveness in uncovering \countAll{} bugs across the Kotlin, Java, Groovy, Scala 2, and Scala 3 compilers. Among the three mutators — \textit{LangShuffler}, \textit{FunctionRemoval}, and \textit{TypeChanger} — \textit{TypeChanger} proves to be the most effective, triggering 11 out of \countAll{} compiler bugs.

\toolName{} also incorporates a manual process to minimize trigger programs and identify the symptoms and root causes of discovered bugs. While 7 bugs can be exposed within a single language, the remaining 17 bugs can only be triggered through cross-language interactions. We further discuss the responsibility of compilers when a language (the user side) references a class/interface from another language (the provider side). The found cross-compilation bug cases show that the user side tends to take the bug responsibilities to adapt to the provider’s type system and inheritance rules.

Currently, \toolName{} only supports class-level and method-level IR. In the next step, we plan to enhance the IR to expression-level and make cross-language programs more versatile and effective. With these IR enhancements, we also aim to design more creative mutators to uncover additional cross-language compilation bugs. In addition, automating the minimization process is a key focus of our future work.

\section{Data-Availability Statement}
The dataset and the source code can be accessed via GitHub~\cite{githublink}. 

